# Ultrasensitive Terahertz Metasurface Biosensor Based on Quasi-Bound States in the Continuum

Junhui Guo[1,†], Bing Dong[2,†], Eryong Zhang[1,†], Qing-An Tu[2,†], Xiaoyong He[1], Xichuan Wu[1], Mingjing Liu[1], Maohua Gong[2], Yan Meng[1, *], Xiang Xi[1, *], Hongcheng Wang[1,*], Zhen Gao[2,*]

[1]*School of Electrical Engineering and Intelligentization, Dongguan University of Technology, Dongguan, 523808, China*
[2]*State Key Laboratory of Optical Fiber and Cable Manufacturing Technology, Department of Electronic and Electrical Engineering, Guangdong Key Laboratory of Integrated Optoelectronics Intellisense, Southern University of Science and Technology, Shenzhen 518055, China.*

[†]*These authors contributed equally to this work.*
[*]E-mail: gaoz@sustech.edu.cn; wanghc@dgut.edu.cn; xix@dgut.edu.cn; mengyan@dgut.edu.cn

The terahertz (THz) spectral regime offers unique opportunities for next-generation biochemical sensing due to its non-destructive, label-free probing capability and strong sensitivity to molecular vibrations. However, conventional THz biosensors remain hampered by intrinsically low-quality factors and limited sensitivity, severely restricting their utility for trace-level biochemical and chemical detection. Here, we report an ultrasensitive THz metasurface biosensor that harnesses quasi-bound states in the continuum (QBICs) with sharp resonances and enhanced light–matter interactions to overcome these limitations. As a proof of concept, the device achieves label-free detection of a sulfur-containing amino acid cysteine, with an ultrahigh sensitivity of 492 GHz/RIU and an ultralow detection limit down to 0.00025 mg/mL. The synergy between QBIC-induced field confinement and meticulous structural optimization of the metasurface underpins this performance, marking a significant advance over conventional THz metasurface biosensing schemes. These results establish QBIC-based metasurfaces as a promising platform for ultrasensitive and high-precision biochemical and chemical sensing, with broad implications for medical diagnostics, food safety, and environmental monitoring.

## 1. Introduction

Metasurfaces, composed of subwavelength meta-atoms, enable flexible manipulation of electromagnetic waves through tailored phase, amplitude, and polarization responses [1–5], providing strong field confinement and enhanced light–matter interactions for biochemical sensing [6–10]. However, conventional terahertz (THz) metasurface sensors are fundamentally limited by radiative and material losses, which restrict their quality ($Q$) factors and sensitivity, particularly in low-concentration or aqueous environments [11–13]. Bound states in the continuum (BICs) offer a fundamental solution to this limitation [14]. As non-radiating eigenmodes embedded within the radiation continuum, BICs can be perfectly localized by symmetry protection or destructive interference [15-20]. In principle, BICs support infinite $Q$ factors and extreme energy confinement [21], enabling breakthroughs in low-threshold lasing, strong nonlinear responses, and ultrasensitive detection [22–27]. Since ideal BICs are inaccessible due to their vanishing radiative coupling, quasi-BICs (QBICs) are engineered via controlled symmetry breaking to introduce weak free-space coupling, yielding measurable ultrahigh-$Q$ resonances with finite linewidths [16–22,27,28]. The degree of symmetry breaking directly governs the radiative leakage and resonance linewidth, providing a practical approach to balance confinement and coupling. Consequently, QBIC-based metasurfaces combine strong near-field enhancement with detectable far-field signatures, enabling highly sensitive detection of subtle refractive-index variations and molecular adsorption events [25,29].

The pursuit of enhanced sensitivity in THz metasurface biosensing [30,31] has driven the development of high-$Q$ QBICs, enabling sharp spectral features and improved detection limits [32]. Building on this principle, phase-sensitive surface-wave strategies have further amplified refractive index responses, while simultaneously enabling deep-learning empowered customized chiral metasurfaces for calibration-free biosensing [33]. The integration of multiple interfering QBIC resonances has subsequently expanded sensing capabilities toward multifunctional platforms capable of real-time monitoring and intelligent discrimination of complex molecular systems [34]. Extending beyond single-resonance operation, gradient and non-periodic metasurface designs have introduced broadband and multi-mode BIC responses, offering compact solutions for wideband THz spectroscopy [35].

In this work, we integrate QBIC physics with THz metasurface biosensing to overcome the sensitivity and linewidth limitations in the THz domain. We design and experimentally demonstrate a THz metasurface biosensor that realizes a tunable quasi-BIC resonance by breaking $C_2$ symmetry between a pair of split-ring resonators. The optimized metasurface supports a high-$Q$ quasi-BIC with pronounced near-field confinement and narrow linewidth, producing large resonance shifts in response to refractive index variations. As a proof of concept, we perform label-free, trace-level detection of cysteine and acetylcysteine, achieving an ultrahigh sensitivity of 492 GHz/RIU and a detection limit down to 0.00025 mg/mL, substantially outperforming conventional THz metasurface sensors. The synergy between QBIC-induced field confinement and rational structural optimization underpins this performance, establishing QBIC-based THz metasurfaces as a versatile platform for high-precision biochemical and chemical sensing in low-concentration environments with potential impact in biochemical and chemical sensing.

## 2. Results and discussion
### 2.1 QBIC biosensor design and optimization

The operation of the proposed THz metasurface biosensor is governed by QBICs, which facilitate strong light–matter interactions and enable ultrahigh sensing sensitivity. The sensing mechanism relies on transducing minute variations in the surrounding refractive index into measurable shifts in the transmission spectrum. As illustrated in Fig. 1(a), analyte types and solution concentrations are represented by droplets of different colors and transparency. When an $x$-polarized THz wave is normally incident along the $z$-axis, the localized electromagnetic field couples to the analyte layer on the metasurface, thereby modifying the effective refractive index of the environment. These refractive-index perturbations induce corresponding resonance shifts in the metasurface spectrum, enabling label-free identification of distinct trace-level analytes and their concentrations. The core of the sensing mechanism lies in the unique capability of the QBIC mode to confine electromagnetic energy within deep subwavelength regions. As shown in the schematic in Fig. 1(b), the electric field is predominantly concentrated at the openings of the split rings and within the narrow coupling channel between them, exhibiting negligible radiation leakage–a hallmark of the QBIC mode. The strong field confinement and suppressed radiation loss jointly enhance the *Q* factor of the system, thereby amplifying its sensitivity to perturbations in the local dielectric environment, enabling trace-level biochemical detection that is challenging to achieve with conventional THz metasurface architectures. The detailed geometry of the metasurface unit cell is illustrated in Fig. 1(c). Two split-ring resonators are symmetrically positioned along the *x*-axis with their openings facing each other. The unit cell is periodic along the *x* and *y* directions with lattice constants $a_1 = 40$ μm and $a_2 = 80$ μm, respectively. The coupling gap between the cavities is $g = 6$ μm; each metallic strip has a width $t = 3$ μm, cavity width $w = 30$ μm, cavity length $h = 28$ μm, and opening widths $d_1 = 2$ μm and $d_2$ (variable). The parameters $d_1$ and $d_2$ that induce structural asymmetry are highlighted in red. The metallic components are modeled as a lossy conductor with an electric conductivity of $\sigma = 4.561 \times 10^7$ S/m to account for Ohmic losses, while the substrate consists of high-purity quartz (dielectric constant $\varepsilon = 3.9$) with a thickness of $d = 500$ μm.

To further elucidate the physical origin of the QBIC mode, we first investigate the symmetric configuration ($d_1 = d_2 = 2$ μm), which supports an ideal symmetry-protected BIC. Numerical simulations are performed using the finite-element method (FEM) with open boundary conditions under normally incident THz illumination along the *z*-axis. As shown in Fig. 1(d), the solved band structure along the high-symmetry directions of the first Brillouin zone (right inset) exhibits a non-radiative BIC point (marked in red) at the Γ point with an eigenfrequency of 2.1 THz. The corresponding electric-field profile (|*E*|, left inset) shows strong confinement within the cavity gaps and coupling channel, confirming the spatially confined nature of the BIC mode. The extracted *Q*-factor distribution along $k_x$ and $k_y$ is shown in Fig. 1(e), exhibiting a pronounced peak at the Γ point. Moreover, the far-field polarization distribution reveals a topological charge of $q = 1$, further confirming the existence of a symmetry-protected BIC.

Although the ideal symmetry-protected BIC possesses an extremely high *Q* factor, its non-radiative nature prevents direct excitation, thereby limiting its sensing applicability. To obtain a measurable resonance, a controlled structural asymmetry is introduced to break the $C_2$ symmetry, converting the BIC into a QBIC with a finite linewidth. The degree of asymmetry is quantified by the parameter $\alpha = (d_2 - d_1)/d_1$, defined as the difference between the open widths of two cavities. To gain further insight into the underlying physical origin of the radiation of the QBIC, a multipole expansion of the far-field scattered power is conducted for the metasurface with *α* = 2 under *x*-polarized illumination, following the electromagnetic multipole expansion theory [36-38]. The scattering contributions from the dominant multipole moments, including the electric dipole (ED), magnetic dipole (MD), electric quadrupole (EQ), magnetic quadrupole (MQ), and toroidal dipole (TD), are evaluated using the expressions below:

$$\text{ED: } \boldsymbol{P} = \frac{1}{i\omega} \int \boldsymbol{j} d^3 r, \tag{1}$$

$$\text{MD: } \boldsymbol{M} = \frac{1}{2c} \int (\boldsymbol{r} \times \boldsymbol{j}) d^3 r, \tag{2}$$

$$\text{EQ: } Q_{\alpha\beta}^{(e)} = \frac{1}{i2\omega} \int \left[ r_\alpha j_\beta + r_\beta j_\alpha - \frac{2}{3}(\boldsymbol{r} \cdot \boldsymbol{j}) \right] d^3 r, \tag{3}$$

$$\text{MQ:} \quad Q_{\alpha\beta}^{(m)} = \frac{1}{3c} \int \left[ (\mathbf{r} \times \mathbf{j})_\alpha r_\beta + (\mathbf{r} \times \mathbf{j})_\beta r_\alpha \right] d^3 r, \tag{4}$$

$$\text{TD:} \quad T = \frac{1}{10c} \int \left[ (\mathbf{r} \cdot \mathbf{j})\mathbf{r} - 2r^2 \mathbf{j} \right] d^3 r, \tag{5}$$

where $\mathbf{j}$ denotes the current density, $\mathbf{r}$ is the position vector, $c$ represents the speed of light in vacuum, and $\omega$ is the angular frequency of the incident wave. By substituting Eqs. (1) – (5) into the FEM framework, the multipole expansions are obtained. As shown in Fig. 1(f), for the configuration with $\alpha = 2$, the radiation is primarily governed by the interference between electric dipole (ED) and magnetic dipole (MD) moments, with the ED component dominating at the QBIC resonance while other higher-order moments remain strongly suppressed. The predominance of the ED moment is further confirmed by the in-plane electric-field distribution ($|E|$) at the Γ point [inset in Fig. 1(f)], which maintains the strong near-field confinement of the BIC mode while exhibiting slightly radiation leakage. This clear evolution from a perfectly non-radiative BIC to an experimentally accessible QBIC highlights the effectiveness of controlled symmetry breaking in engineering high-$Q$, radiatively coupled resonances, thereby enabling ultrasensitive biochemical detection in the THz regime.

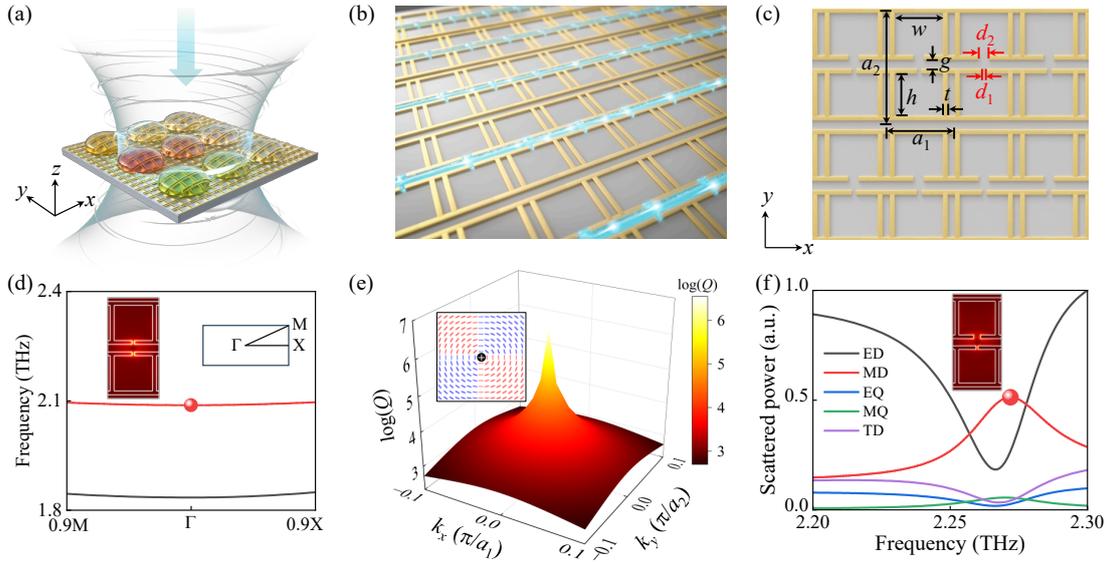

**Fig. 1.** Design concept and physical mechanism of the QBIC metasurface biosensor. (a) The schematic diagram of the terahertz (THz) QBIC metasurface biosensor. Droplets of different colors represent distinct analytes, while transparency indicates concentration (higher transparency corresponds to lower concentration). Variations in refractive index lead to measurable spectral shifts in transmission. (b) The electromagnetic field is primarily localized near the grooves of the metasurface. (c) Composition of the sample: structured gold on a high-purity quartz substrate, with key geometric parameters labeled. (d) Calculated band structure along the first Brillouin zone (right inset) for the symmetric configuration ($d_1 = d_2 = 2$ μm). The non-radiative BIC at the Γ point (red dot) exhibits a strongly localized electric-field distribution ($|E|$, left inset). (e) $Q$-factor distribution (colormap) for the BIC mode, the corresponding far-field polarization distributions are shown in the inset, featuring a $q = +1$ topological charge of the Γ-point BIC. (f) Multipole expansion of the far-field scattered power for the asymmetric (QBIC) configuration ($d_1 = 2$ and $d_2 = 6$ μm). Radiation is dominated

by the electric dipole (ED) moment, as verified by the in-plane electric-field distribution ($|E|$) at the QBIC resonance (red dot).

To optimize the performance of the QBIC-based biosensor, we numerically calculate the transmission spectra as a function of the asymmetry parameter $\alpha$. As the results shown in Fig. 2(a), when $\alpha = 0$, the configuration preserves $C_2$ symmetry, resulting in extremely low transmission within the investigated frequency regime, consistent with the non-radiative nature of the BIC mode. When $\alpha$ deviates from zero, the $C_2$ symmetry is broken and a bright transmission band emerges, implying a measurable coupling to radiative channels. To illustrate the continuous evolution from BIC to QBIC, we plot the transmission spectra for $\alpha = -0.5, 0, 0.5, 2$, and 3 in Fig. 2(b), it can be observed that, a clear transmission peak appears when $\alpha$ gradually deviates from zero, and both the linewidth and peak amplitude increase with $|\alpha|$, implying the enhanced radiation leakage of QBICs.

The leaky characteristics of the QBICs are quantitively evaluated using the Fano formula expressed as [39]:

$$T_{Fano} = t_0 \times \left| a_1 + ia_2 + \frac{b}{\omega - \omega_0 + i\gamma} \right|^2, \qquad (6)$$

where $\omega_0$ and $\gamma$ denote the resonance frequency and the damping rate, respectively, while $a_1$ and $a_2$ are fitting constants. The quality factor is defined as $Q = \omega_0/2\gamma$. The extracted $Q$ values for different $\alpha$ are plotted by the red curve in Fig. 2(c). The results reveal that radiation loss and intrinsic Ohmic dissipation of the gold film jointly induce an obvious suppression on the $Q$ factor. Moreover, the quality factor $Q$ decreases with increasing asymmetric parameter $\alpha$, following an inverse-square dependence, $Q \propto \alpha^{-2}$ [25]. Besides the $Q$ factor, the transmission peak intensity $I = |T_2 - T_1|$ is introduced, where $T_2$ corresponds to the peak value and $T_1$ corresponds to the baseline value of the transmission spectrum [see inset of Fig. 2(c)]. As the black line depicted in Fig. 2(c), the intensity $I$ increases monotonically with $\alpha$, implying an improved peak-to-trough contrast of the transmission spectrum. To balance the $Q$ factor and the intensity, we define a figure of merit (FoM) as [40]:

$$\text{FoM} = Q \times I. \qquad (7)$$

The calculated FoM is shown in Fig. 2(d), where a maximum value of FoM = 25 appears at $\alpha = 2$, indicating that this configuration achieves the optimal trade-off between high $Q$ and sufficient radiative coupling. Thus, $\alpha = 2$ is selected as the optimized design for subsequent sensing studies.

To evaluate the sensing performance of the optimized THz metasurface biosensor, we further simulated the frequency shift ($\Delta f$) of the QBIC resonance under analyte films with varying refractive indexes and thicknesses. The thickness range of 1–20 μm was chosen based on the simulated electric field decay, which shows that the field intensity decreases to below 1% of its maximum value at 5 μm, indicating negligible analyte contribution beyond this distance. This range also covers the estimated experimental thicknesses. The refractive index range of 1.0–2.0 was selected to encompass typical values for dry biomolecular films (1.4–1.6) and to clearly illustrate the linear dependence of $\Delta f$ on the refractive index. The results are shown in Fig. 2(e), for a fixed thickness, $\Delta f$ increases linearly with the refractive index. Conversely, for a fixed refractive index, increasing the analyte thickness leads to a gradual shift and saturation of $\Delta f$. This saturation behavior is further illustrated in Fig. 2(f), when the thickness exceeds 5 μm, the frequency shift begins to enter saturation, and as it approaches 20 μm, a saturation regime is observed, the simulated results (colored dots) are well fitted by a function of $\Delta f = At/(1+t)$, with A being a coefficient that depends on the refractive index and t being the thickness of the analyte. The saturation originates from the spatial distribution of the electromagnetic field, which leads to the upper portion of the analyte lying outside the near-field region (beyond approximately 20 μm) and no longer contributing effectively to the resonance shift. The sensing sensitivity S is defined as $S = \Delta f/\Delta n$. Under saturated thickness conditions, the biosensor achieves a sensitivity of up to 492 GHz/RIU, while maintaining a detection sensitivity of 329 GHz/RIU even with ultra-thin analyte coverage (1 μm thickness). With this paradigm, it is concluded that the biosensor is quite sensitive to changes in refractive index for a fixed analyte thickness. Consequently, coating the metasurface with thin analyte layers of varying concentrations yields distinct and measurable resonance shifts, establishing a theoretical foundation for trace-level biochemical detection.

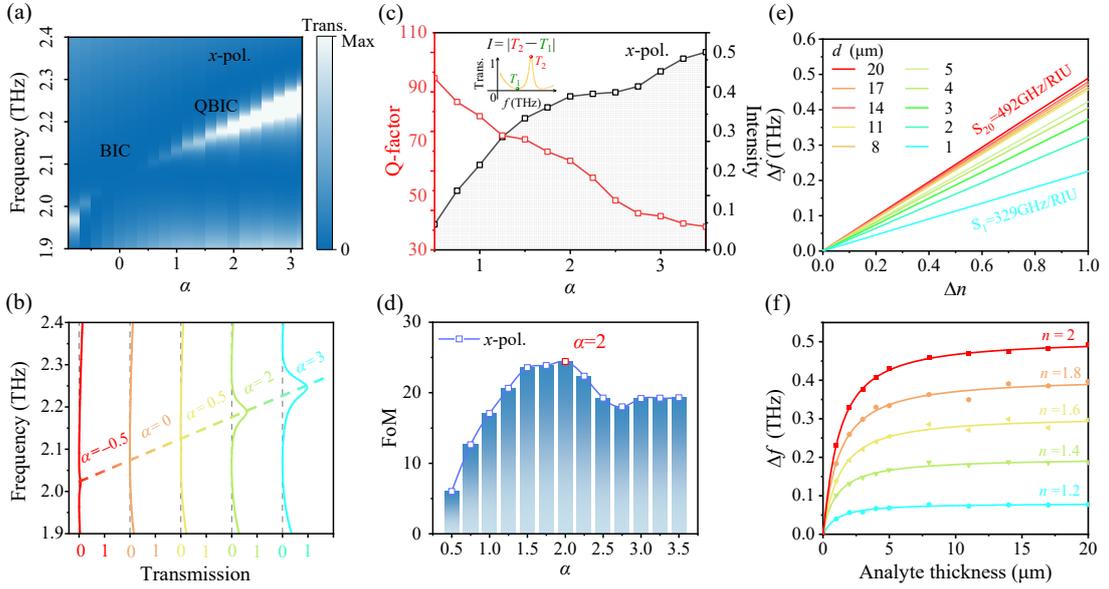

**Fig. 2.** Optimization and sensing performance of the QBIC metasurface biosensor. (a) Simulated transmission spectra as a function of the asymmetric parameter $\alpha$. (b) Evolution of the transmission spectra for selected values of $\alpha = -0.5, 0, 0.5, 2$, and 3, illustrating the transition from BICs to QBICs. (c) Extracted $Q$ factors (red line) and the transmission intensity $I$ (black curve) under x-polarized illumination. Inset: the definition of $T_1$ and $T_2$. (d) Calculated figure of merit (FoM = $Q \times I$) as a function of $\alpha$, reaching a maximum value at $\alpha = 2$. (e) Simulated frequency shift ($\Delta f$) as a function of analyte refractive index and thickness. (f) The gradual evolution of the frequency shift $\Delta f$ for analytes with fixed refractive indexes and increasing thickness. Symbols: simulated results. Lines: fitting curves.

**2.2 Sample preparation and experimental measurement**

Cysteine ($C_3H_7NO_2S$) and acetylcysteine ($C_5H_9NO_3S$) are sulfur-containing amino acid derivatives with critical biological roles: cysteine is essential for glutathione synthesis and antioxidant defense, while acetylcysteine is a clinically used mucolytic and antidote. Both molecules contain polar functional groups (thiol, amino, carboxyl) whose vibrational and rotational modes lie in the THz fingerprint region, enabling direct, label-free detection via THz spectroscopy. The detection and measurement process follows the workflow illustrated in Fig. 3(a). Biological powder samples of acetylcysteine and cysteine, both purchased from Aladdin Biochemical Co., Ltd. (Shanghai, China), were first weighed using a precision electronic balance (accuracy: 0.001 g). To evaluate the sensing performance, the quantified powder samples were dissolved in deionized water to prepare aqueous solutions with controlled concentrations ranging from 0.00025 mg/mL to 40 mg/mL, along with a blank reference for comparison. After complete dissolution, each solution is labeled and stored for subsequent measurements. Using a micropipette, 20 µL of each solution is vertically titrated onto the metasurface sensing area. To avoid the confounding effects of water absorption in the THz band, the samples are dried before THz measurements. After titration, the samples are dried in an oven at 50–60 °C for 10 minutes. This gentle heating accelerates evaporation while avoiding thermal degradation of the biomolecules, forming a uniform thin film on the biosensors.

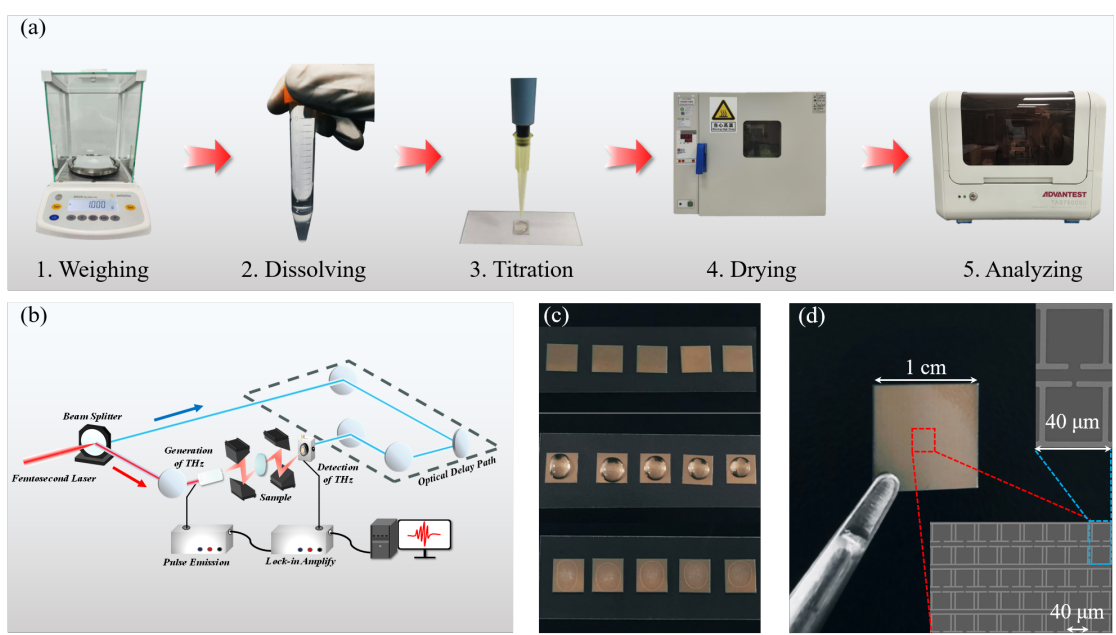

**Fig. 3.** Experimental workflow and terahertz time-domain spectroscopy (THz-TDS) characterization of the QBIC metasurface biosensor. (a) Workflow for sample preparation, including weighing, dissolving, titration, drying, and analyzing. (b) Schematic of the THz-TDS set-up. Ultrafast laser pulses generate and detect broadband THz radiation transmitted through the sample. (c) Photographs of the fabricated biosensor arrays: pristine (top), coated with analyte solutions of different concentrations before drying (middle), and after drying (bottom). (d) Optical image of the fabricated metasurface biosensor and its geometric details (insets). The chip size is approximately 1 cm × 1 cm, with a unit cell dimension of 40 μm × 80 μm.

The optical response of the prepared biosensors is characterized using a THz time-domain spectroscopy (THz-TDS) system, schematically shown in Fig. 3(b). The experimental measurements are performed with a commercial transmission-type spectrometer (TAS7500SU, Advantest, Japan). The system consists of a femtosecond laser source, a THz emitter, a detector, an optical delay line, and a sample stage. Ultrafast laser pulses are divided into two paths: one portion of the laser beam excites the emitter to generate broadband THz radiation, which is transmitted through the sample and subsequently detected, while the other portion is routed through a variable delay path to temporally scan the THz pulse. By recording the transmitted time-domain signal and applying fast Fourier transformation (FFT), both amplitude and phase spectra are obtained simultaneously, allowing the extraction of frequency-dependent optical parameters such as transmission, absorption, and reflection. This setup configuration enables accurate and non-destructive probing of the dielectric response of biochemical samples in the THz regime.

Representative fabricated samples coated with acetylcysteine and cysteine solutions of various concentrations are illustrated in Fig. 3(c). The upper panel shows the pristine biosensor arrays before sample titration, while the middle and lower panels correspond to the states before and after drying, respectively. For each concentration, 20 μL of the prepared aqueous solution is precisely titrated onto the sensing area using a micropipette. During the drying stage, the solvent gradually evaporated, resulting in the uniform and adherent thin films on the biosensor surface. This ensured consistent surface morphology and uniform analyte distribution, which are essential for achieving reliable and repeatable THz measurements. The structural pattern and geometric dimensions of the biosensor are presented in Fig. 3(d). Each chip measures approximately 1 cm × 1 cm and contains a periodic microstructured gold array on a quartz substrate, with unit cell dimensions of 40 μm (*x* direction) and 80 μm (*y* direction), respectively. Such precisely fabricated geometry supports strong THz field confinement and resonant enhancement, thereby improving the sensitivity of the device to minute dielectric variations induced by biomolecular adsorption. This microstructured configuration provides an optimized platform for ultrasensitive and quantitative biochemical detection in the THz frequency regime.

## 2.3 Detection results and analysis

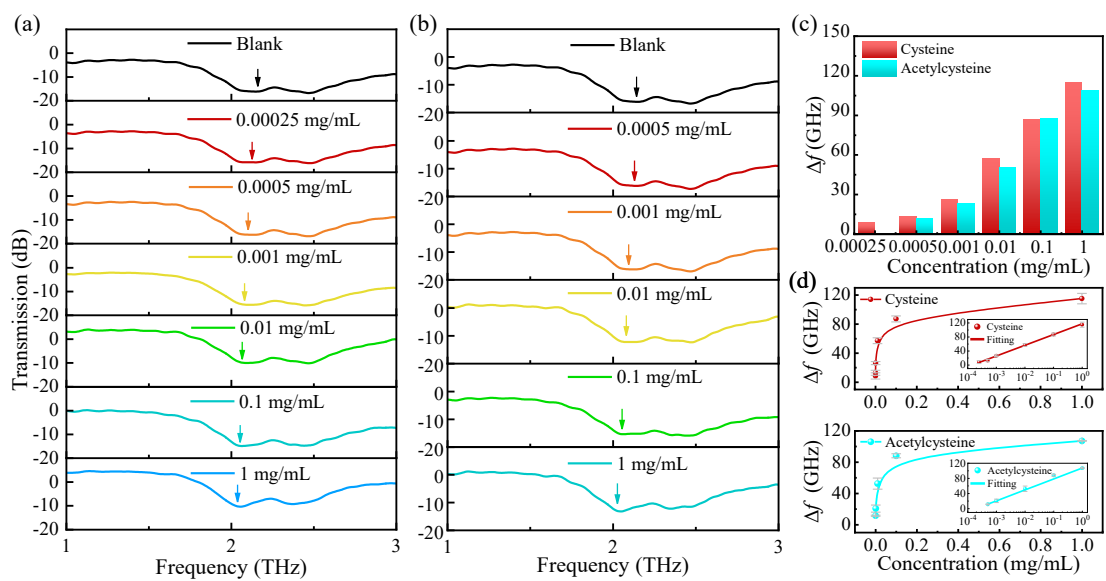

**Fig. 4.** Detection results and analysis in the low-concentration regime (0–1 mg/mL). (a)-(b) Measured transmission spectra of the metasurface biosensor coated with acetylcysteine (a) and cysteine (b) solutions of different concentrations. (c) Comparison of the resonance frequency shifts ($\Delta f$) for cysteine (red) and acetylcysteine (cyan) solutions of different concentrations. (d) Measured resonance shifts (symbols) and logarithmic fitting curves (solid lines in the inset) for cysteine (upper panel) and acetylcysteine (lower panel), showing a clear logarithmic

dependence of Δ*f* on concentration. The error bars (gray symbols) indicate the deviation from the mean value based on nine repeated measurements.

The sensing performance of the QBIC-based THz metasurface biosensor is evaluated using THz-TDS technology across the frequency range of 1–3 THz. As shown in the first panel of Figs. 4(a) and 4(b), the transmission spectrum of the reference (blank) sample exhibits two transmission dips (corresponding to two absorption peaks), a dominated one at $f_0 \approx 2.1$ THz corresponding to the QBIC mode of interest, while the secondary one around 2.5 THz corresponding to a higher order mode, which is not considered in this work. For quantitative analysis, the measurements are divided into two regimes: the low-concentration regime (0–1 mg/mL) and the high concentration (1–40 mg/mL) regime. As shown in Figs. 4(a) and 4(b), in the low-concentration regime, when the concentrations of cysteine and acetylcysteine increase, the QBIC resonance gradually redshifts from $f_0 \approx 2.1$ THz to $f_0 \approx 2.0$ THz. The minimum resolvable concentration is determined by the spectral resolution of the THz-TDS system ($\approx 7.6$ GHz), corresponding to detection limits of 0.00025 mg/mL for cysteine and 0.0005 mg/mL for acetylcysteine, representing the sensor's limit of detection.

A comparison of the frequency shifts ($\Delta f$) induced by the two analytes is presented in Fig. 4(c), revealing that cysteine generally induces a larger redshift than acetylcysteine in the low-concentration regime. Fig. 4(d) shows the measured $\Delta f$ as a function of concentration, which can be perfectly fitted (with all coefficients of determination satisfying $R^2 > 0.99$ throughout this work) using a logarithmic function. For cysteine, the fitting yields $\Delta f = 30.6 \log_{10}(C_C) + 117.5$ (inset, upper panel), demonstrating an excellent logarithmic relationship between the frequency shift and analyte concentration [41]. The corresponding results for acetylcysteine are shown in the lower panel of Fig. 4(d) and can be fitted by $\Delta f = 28.9 \log_{10}(C_A) + 107.45$ (inset, lower panel). The larger slope obtained for cysteine further confirms the higher sensitivity of the biosensor to cysteine molecules in the low-concentration regime.

Similarly, the detection results for cysteine and acetylcysteine at higher concentrations (1–40 mg/mL) are shown in Figs. 5(a) and 5(b), respectively. As the concentration increases, the transmission dips continue to redshift, reaching frequencies as low as 1.7 THz. A comparison between the two analytes in Fig. 5(c) reveals that, at higher analyte loadings, acetylcysteine induces a larger redshift than cysteine, indicating a concentration-dependent reversal of sensitivity between the two analyte solutions around the critical concentration of approximately 1 mg/mL. As shown in Fig. 5(d), in the high-concentration regime, the frequency shift exhibits a linear rather than a logarithmic dependence on concentration. The fitted linear relationships are $\Delta f = 7.76 C_A + 93.38$ and $\Delta f = 8.00 C_A + 99.96$ for cysteine and acetylcysteine, respectively. The larger slope obtained for acetylcysteine confirms its higher sensitivity compared with cysteine in the high-concentration regime. The distinct fitting functions reflect a fundamental shift in sensing mechanism: at low concentrations (<1 mg/mL), the logarithmic response arises from surface adsorption on high-field hot spots (sub-monolayer regime); at high concentrations (>1 mg/mL), the linear response originates from the formation of a quasi-continuous film whose thickness scales with concentration. The transition at ~1 mg/mL marks the crossover from adsorption-dominated to film-thickness-dominated sensing.

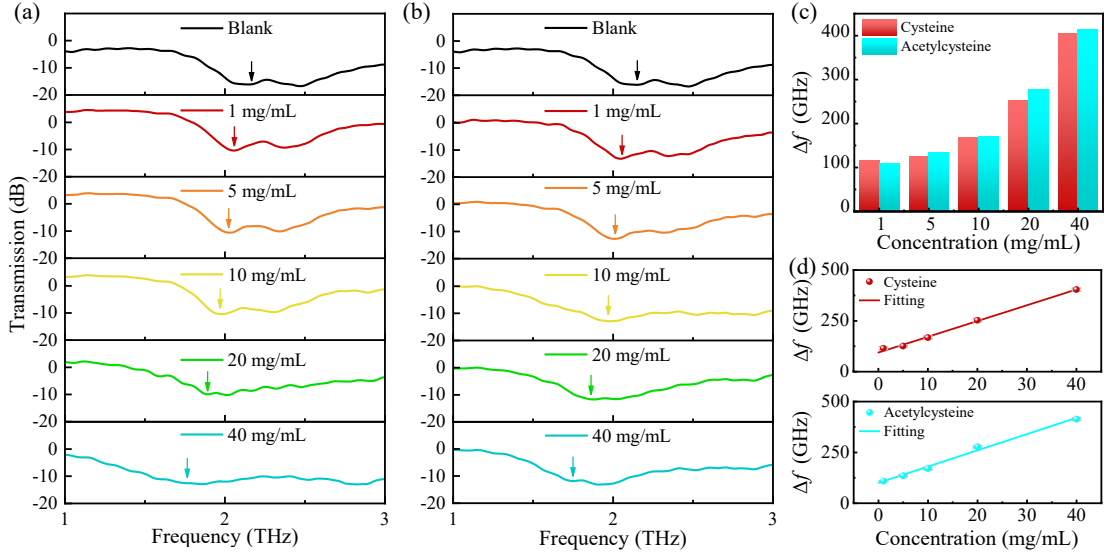

**Fig. 5.** Detection results and analysis in the high-concentration regime (1–40 mg/mL). (a)-(b) Measured transmission spectra of the metasurface biosensor coated with acetylcysteine (a) and cysteine (b) of different concentrations. (c) Comparison of the resonance frequency shifts ($\Delta f$) for cysteine (red) and acetylcysteine (cyan) solutions of different concentrations. (d) The measured results (dots) and linear fitting (lines) of $\Delta f$ versus concentration for cysteine (upper panel) and acetylcysteine (lower panel). The error bars (gray symbols) indicate the deviation from the mean value based on nine repeated measurements.

Compared to other sensing approaches, the proposed THz metasurface biosensor achieves an ultralow detection limit of 0.00025 mg/mL, which is two orders of magnitude lower than recently reported THz sensors (0.030 mg/mL in [42] and 0.063 mg/mL in [43]), enabling trace-level biochemical identification that surpasses most conventional THz or plasmonic sensors. Benefiting from the high-$Q$ QBIC resonance and strong near-field enhancement, the system requires only minute analyte quantities (~ 5 ng for cysteine), significantly reducing material consumption while maintaining high accuracy. Furthermore, the sensing process is remarkably efficient–a single spectral scan is completed within several tens of seconds, and the entire workflow, including sample pretreatment and drying, can be finished in approximately ten minutes. To provide a comprehensive comparison, the performance of our sensor is summarized alongside other reported methods for cysteine-related detection in Table 1. As shown, our device achieves the lowest detection limit (0.00025 mg/mL) among the compared techniques, while offering additional advantages including simplified experimental procedures, minimal sample consumption, and rapid measurement. This rapid response, combined with high sensitivity and minimal sample requirement, demonstrates the potential of the QBIC-based THz metasurface biosensor as a practical, high-throughput platform for real-time biochemical and environmental detection.

| Measured Method | Biological Samples | Detection limit (mg/mL) | Reference |
|---|---|---|---|
| Electrochemiluminescence | L-Cysteine | 0.030 | [42] |

| Paper/Cloth-Based Microfluidic Chip | L-Cysteine | 0.063 | [43] |
| THz Biosensor (QBIC) | Homocysteine | 0.00169 | [31] |
| This Work (QBIC Metasurface) | L-Cysteine, NAC | 0.00025 | This work |

**Table 1** Performance comparison of the proposed QBIC metasurface sensor with other reported detection methods.

## 3. Conclusion

In conclusion, we have demonstrated an ultrasensitive terahertz metasurface biosensor based on quasi-bound states in the continuum (QBICs) for label-free biochemical detection. By introducing controlled structural asymmetry, the device supports radiatively accessible QBIC modes with strong field confinement and enhanced light–matter coupling. The biosensor enables quantitative detection of cysteine and acetylcysteine across a broad concentration range, exhibiting distinct logarithmic and linear responses in low- and high-concentration regimes, respectively. Remarkably, the detection limit reaches as low as 0.00025 mg/mL, corresponding to only 5 ng of analyte, highlighting its capability for trace-level biochemical sensing. The metasurface design features simple fabrication, fast readout, and excellent repeatability, offering a practical and scalable solution for THz biosensing. These results position the QBIC-based platform as a versatile tool for rapid, non-destructive detection of biomolecules, with promising applications in medical diagnostics, environmental monitoring, and food safety.

**Acknowledgments**
Y. M. acknowledges the support from the National Natural Science Foundation of China under Grant No. 12304484, Guangdong Basic and Applied Basic Research Foundation under grant No. 2024A1515011371. Z.G. acknowledges funding from the Key Research and Development Program of the Ministry of Science and Technology (grant no. 2025YFA1412300), National Natural Science Foundation of China (grant no. 62361166627 and 62375118), Guangdong Basic and Applied Basic Research Foundation (grant no. 2024A1515012770), Shenzhen Science and Technology Innovation Commission (grants no. 202308073000209), and High-level Special Funds (grant no. G03034K004). X. X. acknowledges the support from the National Natural Science Foundation of China under grant No.62405053, Basic and Applied Basic Research Foundation of Guangdong Province under grant No. 2025A1515012229. H. W. acknowledges the support from the National Key R&D Program of China (Grant No. 2023YFC2206500)

**Conflict of Interest**
The authors declare no conflicts of interest.

**Data Availability Statement**
The data that support the findings of this study are available from the corresponding author upon reasonable request.

**Keywords**

Quasi–bound states in the continuum (QBICs), Terahertz metasurface biosensor, Trace-level detection